\documentclass[conference]{IEEEtran}
\IEEEoverridecommandlockouts
\usepackage{cite}
\usepackage{amsmath,amssymb,amsfonts}
\usepackage{graphicx}
\usepackage{textcomp}
\usepackage{xcolor}
\usepackage{subcaption}
\usepackage{xcolor}
\usepackage[numbers,sort&compress]{natbib}
\def\BibTeX{{\rm B\kern-.05em{\sc i\kern-.025em b}\kern-.08em
    T\kern-.1667em\lower.7ex\hbox{E}\kern-.125emX}}

\makeatletter 
\newcommand{\linebreakand}{%
  \end{@IEEEauthorhalign}
  \hfill\mbox{}\par
  \mbox{}\hfill\begin{@IEEEauthorhalign}
}
\makeatother 

\begin{document}

\title{Spatio-Temporal Characterization of Qubit Routing in Connectivity-Constrained Quantum Processors
 \thanks{Authors gratefully acknowledge funding from the European Commission through HORIZON-EIC-2022-PATHFINDEROPEN-01-101099697 (QUADRATURE) and grant HORIZON-ERC-2021-101042080 (WINC).}
 }

\author{\IEEEauthorblockN{Sahar Ben Rached}
\IEEEauthorblockA{\textit{NanoNetworking Center in Catalunya} \\
\textit{Universitat Polit\`ecnica de Catalunya}\\
Barcelona, Spain \\
sahar.benrached@upc.edu}
\and
\IEEEauthorblockN{Carmen G. Almud\'ever}
\IEEEauthorblockA{\textit{Computer Engineering Department} \\
\textit{Universitat Polit\`ecnica de Val\`encia}\\
Val\`encia, Spain \\
cargara2@disca.upv.es}
\and
\IEEEauthorblockN{Eduard Alarc\'on}
\IEEEauthorblockA{\textit{NanoNetworking Center in Catalunya} \\
\textit{Universitat Polit\`ecnica de Catalunya}\\
Barcelona, Spain \\
eduard.alarcon@upc.edu}
\and

\linebreakand

\IEEEauthorblockN{Sergi Abadal}
\IEEEauthorblockA{\textit{NanoNetworking Center in Catalunya} \\
\textit{Universitat Polit\`ecnica de Catalunya}\\
Barcelona, Spain \\
abadal@ac.upc.edu}

}

\maketitle

\begin{abstract}
Designing efficient quantum processor topologies is pivotal for advancing scalable quantum computing architectures. The communication overhead, a critical factor affecting the execution fidelity of quantum circuits, arises from inevitable qubit routing that brings interacting qubits into physical proximity by the means of serial SWAP gates to enable the direct two-qubit gate application. Characterizing the qubit movement across the processor is crucial for tailoring techniques for minimizing the SWAP gates. This work presents a comparative analysis of the resulting communication overhead among three processor topologies: star, heavy-hexagon lattice, and square lattice topologies, according to performance metrics of communication-to-computation ratio, mean qubit hotspotness, and temporal burstiness, showcasing that the square lattice layout is favourable for quantum computer architectures at a scale.
\end{abstract}
\section{Introduction}
Quantum computing technology promises a high computation power to solve complex problems that are intractable with classical supercomputers in certain application fields. Yet, the challenge of building Quantum Processing Units (QPUs) that robustly satisfy technological expectations persists, despite significant advancements in the field in recent years. Among the pressing challenges is the development of efficient compilation techniques that take account of the limited connectivity of quantum processors and minimize the resulting qubit routing overhead. Indeed, qubits in contemporary quantum devices \cite{b1} are arranged in topologies with limited connectivity between them where only nearest-neighbouring qubit interactions are applicable. Compiling and executing high-level quantum circuits on such architectures requires adding SWAP gates to bring interacting qubits that are not initially placed in adjacent positions on the processor into physical proximity to enable the direct application of two-qubit gates \cite{b2}. For circuits of large sizes and high density of two-qubit gates, the extensive application of SWAP gates to transfer the quantum states between qubits results in high gate count and circuit depth \cite{b3}, thus restricting the implementation of complex quantum algorithms reliably. 

Several methods have been proposed to reduce the number of SWAPs in the circuit compilation process by optimizing the qubit mapping from logical to physical qubits and their routing across the chip (e.g. choosing the shortest path) \cite{b4}-\cite{b7}. Most fundamentally, advancing the hardware of quantum computing systems to support more interconnected qubit layouts would inherently reduce the need for SWAP gates. On the other hand, a densely interconnected processor topology leads to prohibitive crosstalk and adds to the overall complexity and cost of the system \cite{b8}. As quantum computing technology has evolved from its nascent stages, the processor architecture progressed from the initial prototypes featuring one-dimensional (1D) linear qubit arrays \cite{b9} to the more complex two-dimensional (2D) configurations \cite{b10}-\cite{b12}. Yet, the development and optimization of processor topologies remains an active field of research \cite{b13,b14}.

The standard metrics for assessing the qubit routing overhead are static parameters, mainly the number of added SWAP gates, the circuit depth and execution time \cite{b15}. For our work, we analyze the communication overhead and qubit routing patterns in monolithic processors of various topologies to conduct a spatio-temporal characterization of the qubit movement as imposed by the compilation process. We interpret the communication overhead according to three metrics: the communication-to-computation operations ratio, the mean qubit hotspotness and the system burstiness. To this aim, we compile quantum circuits of different structures on three topologies: star, heavy-hexagon (heavy-hex) lattice, and square lattice. The heavy-hex \cite{b16} and square lattice \cite{b17} architectures are currently utilized in the construction of real quantum processors, indicating their practical applicability and technological viability. The star topology is mainly adopted in the prototypical development stages of quantum chips \cite{b18}. Its efficiency has been demonstrated for certain simulations, as detailed in \cite{b19}, highlighting its potential in specialized computational tasks. Evaluating the spatio-temporal qubit movement in quantum circuit execution resulting from the underlying topology aims to identify the architectural bottlnecks and guide the advancement of compilation techniques that better manage the communication load. 
\section{Methodology and experiments}
\begin{figure}[htbp]
\centerline{\includegraphics[scale=0.35]{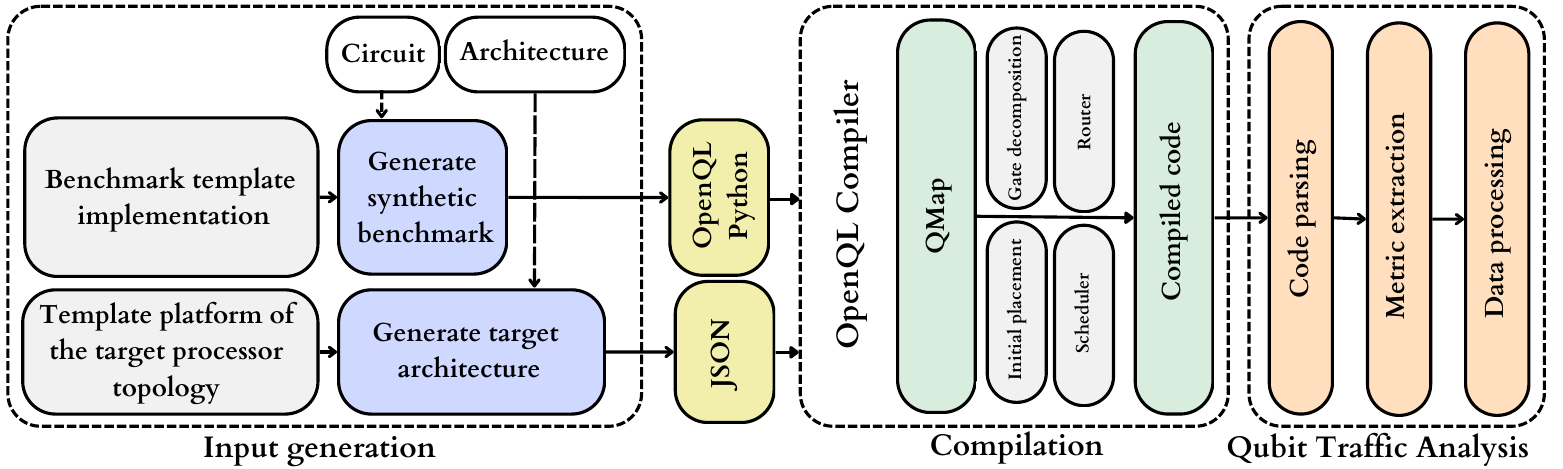}}
\caption{Flow diagram of the qubit routing analysis software tool}
\label{methodology}
\end{figure}

For our work, we select four quantum circuits of different logical structures: the Cuccaro Adder (Cuccaro) \cite{b20}, the Grover's main routine (Grover) \cite{b21}, the Quantum Approximate Optimization Algorithm \cite{b22} solving the MaxCut problem with a random Erdos Renyi \cite{b23} input graph of edge probability 0.2 (QAOA\_02), and the Quantum Fourier Transform (QFT) \cite{b24}. We compile the circuits of increasing sizes considering the aforementioned processor topologies. We use OpenQL \cite{b25} to optimize and map the quantum circuits to the underlying layout. By extracting data on qubit movements throughout the program execution, we generate and analyze the communication traces according to the formulated metrics, which are further processed and graphically visualized, as depicted in Fig. \ref{methodology}.
We define the following metrics to represent the communication overhead associated with each processor topology:
\begin{enumerate}
    \item \textbf{Communication-to-Computation Ratio (CCR)}: defines a normalized quantity of the amount of communication operations applied over runtime. We interpret the communication-to-computation ratio as the ratio of SWAP operations applied to transfer the quantum states $n_{SWAP}$ to the computation gates applied to execute the input algorithm $n_{ops}$:
    \begin{equation}
    \mathbf{CCR} = \dfrac{n_{SWAP}}{n_{ops}} \label{ccr-eq}
    \end{equation}

    \item \textbf{Mean Qubit Hotspotness}: evaluates the imbalance of communication operations across qubits which might degrade the overall computational fidelity if it is large. We represent the mean qubit hotspotness as the ratio of the variance of SWAP operations applied to physical qubits $\sigma^2_{SWAP}$ to the mean of SWAP operations applied to physical qubits $\mu_{SWAP}$ over runtime: 
    \begin{equation}
    \overline{H}_{qubit} = \dfrac{\sigma^2_{SWAP}}{\mu_{SWAP}} \label{hotspotness-eq}
    \end{equation}

    \item \textbf{Temporal Burstiness}: estimates the amount of communication operations occurring in parallel, possibly generating routing contention. Accordingly, we monitor the SWAP operations clustered in well-defined timeslices over the execution time. The system is supposed bursty if there exists timeslices exhibiting a high number of SWAP operations clustered subsequently, followed by timeslices of low or spaced SWAP gates. We characterize the burstiness of a quantum system as the ratio of the variance of parallel SWAP operations $\sigma^2_{SWAP(t)}$ to the mean of parallel SWAP operations throughout the execution timeslices $\mu_{SWAP(t)}$:
    \begin{equation}
    \overline{B} = \dfrac{\sigma^2_{SWAP(t)}}{\mu_{SWAP(t)}}
    \end{equation}    
\end{enumerate}
\section{Results}
\begin{figure*}
\centering
\begin{subfigure}{.25\textwidth}
  \centering
  \includegraphics[width=1\textwidth]{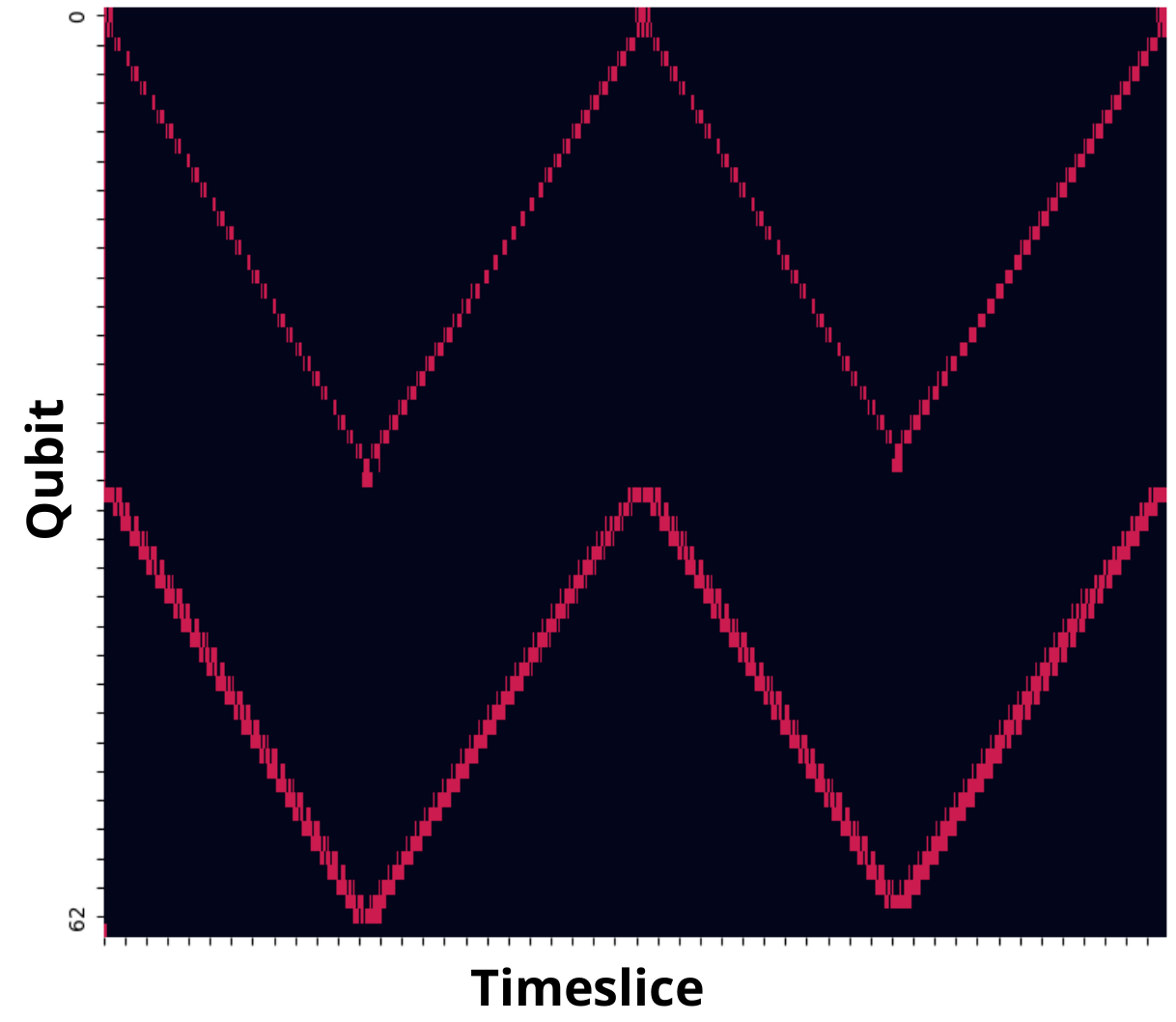}
  \caption{Virtual mapping}
  \label{traces:sub1}
\end{subfigure}%
\begin{subfigure}{.25\textwidth}
  \centering
  \includegraphics[width=1\textwidth]{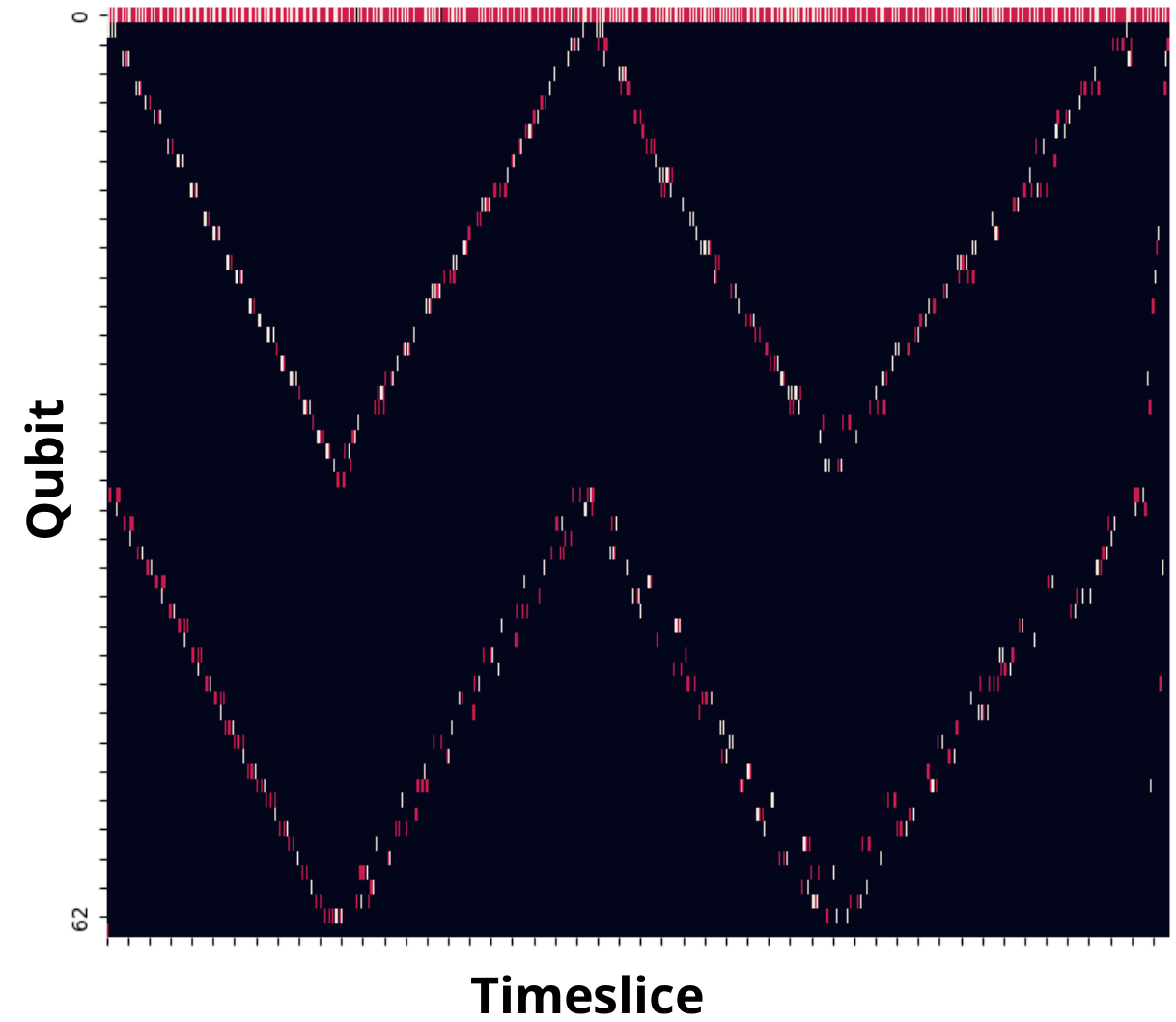}
  \caption{Star topology}
  \label{traces:sub2}
\end{subfigure}%
\begin{subfigure}{.25\textwidth}
  \centering
  \includegraphics[width=1\textwidth]{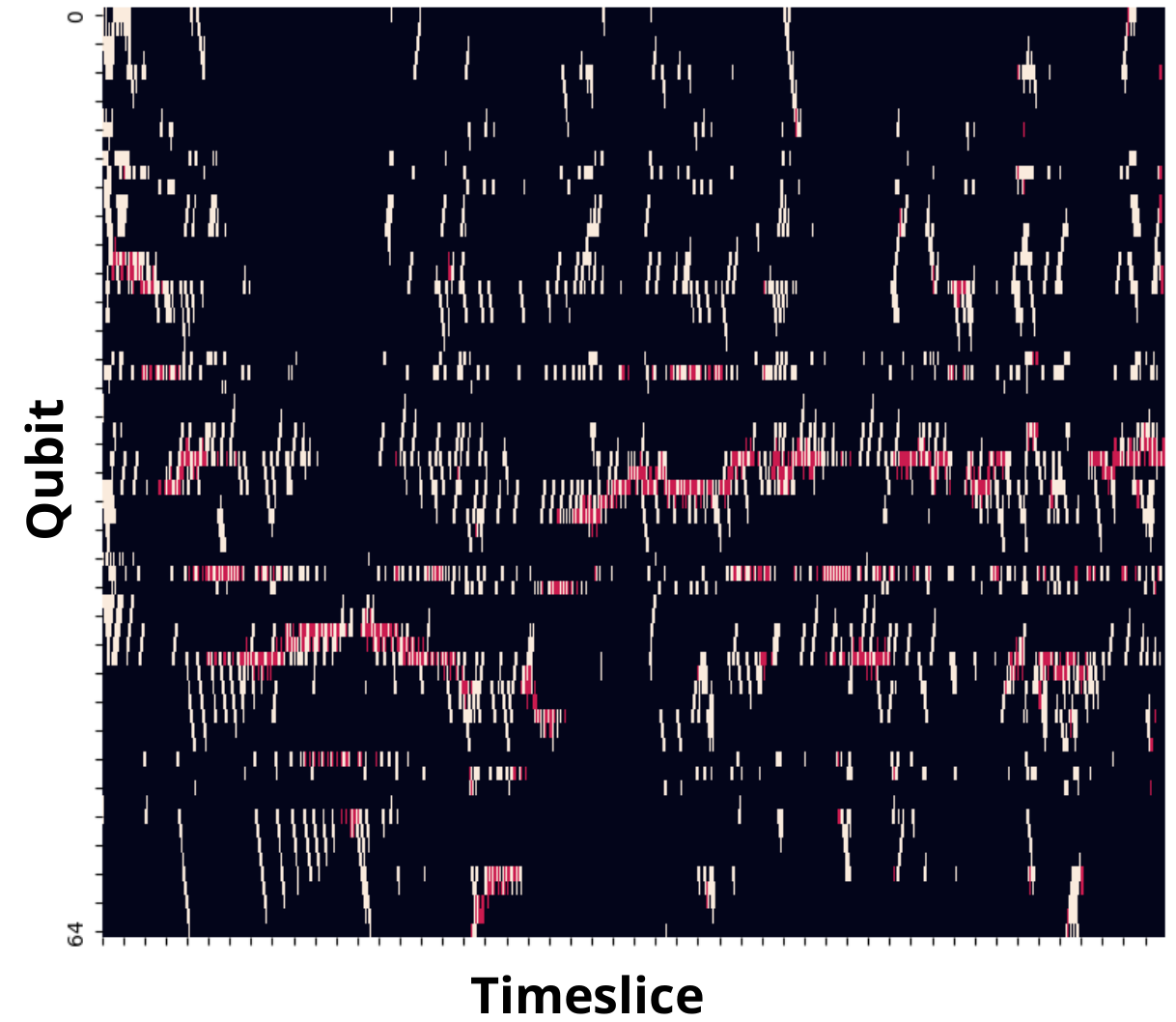}
  \caption{Heavy-hex lattice topology}
  \label{traces:sub3}
\end{subfigure}%
\begin{subfigure}{.25\textwidth}
  \centering
  \includegraphics[width=1\textwidth]{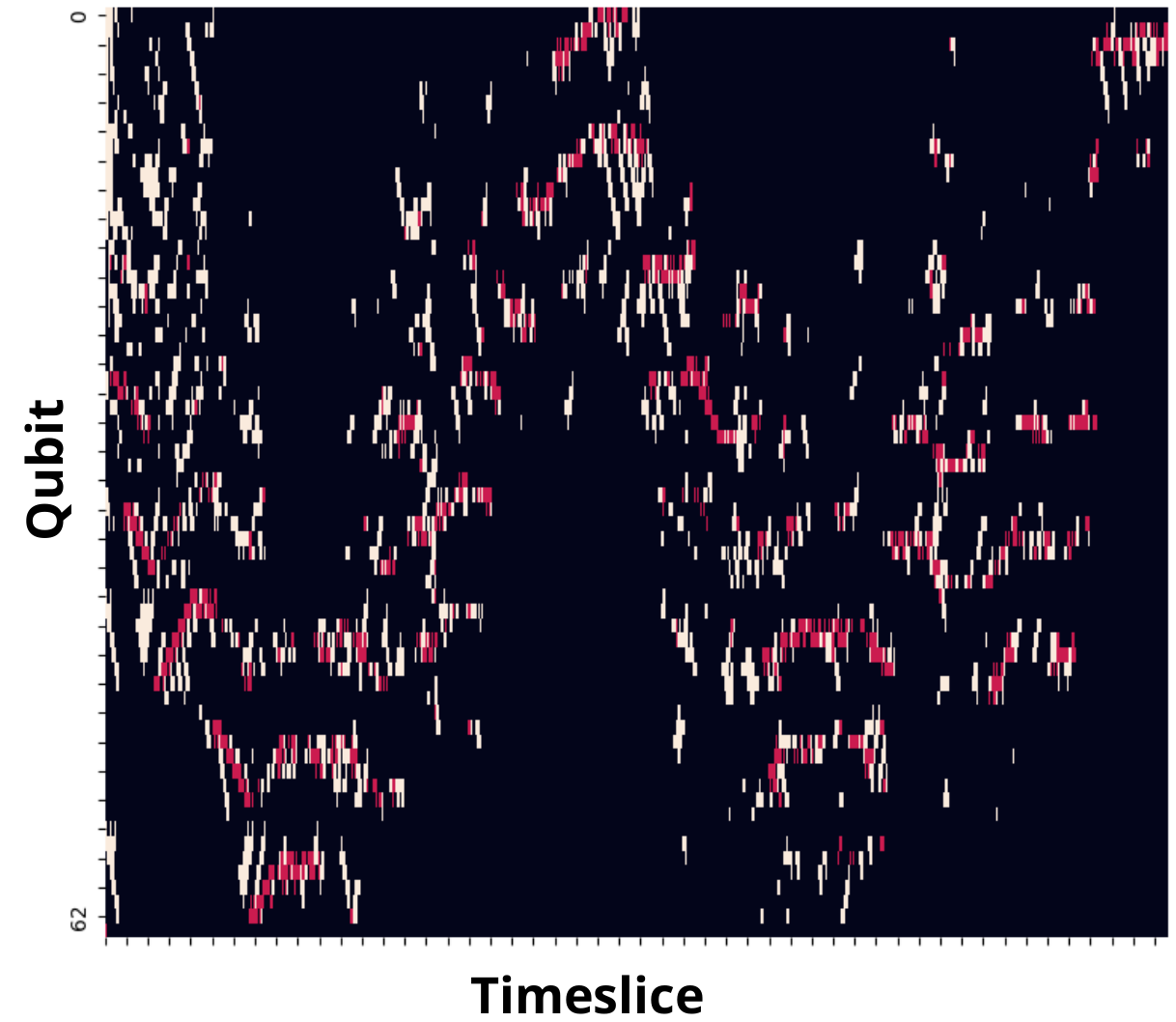}
  \caption{Square lattice topology}
  \label{traces:sub4}
\end{subfigure}%
\caption{Grover's circuit mapping to virtual qubits (a), and physical qubits (b), (c), (d). Computation operations, SWAP gates, and idling times are represented in red, white and black, respectively.}\label{traces}
\end{figure*}
In our examination of the Grover circuit compilation, we present a comparative visualization between the circuit mapping to virtual qubits and to physical qubits across the three different topologies in Fig. \ref{traces}. The virtual mapping showcases the gate application sequence and the involved qubits regardless of the physical constraints of the quantum processor. On the other hand, the physical mapping requires a series of modifications to comply with the specific interconnectivity of each topology, which consist of adding SWAP gates. Mapping the circuit to the star topology tends to preserve much of the original circuit logical structure, given that the central qubit's direct links to all others limits the number of SWAP gates. In contrast, mapping to the heavy-hex and square lattice topologies demands significant restructuring, due to their qubit arrangements being less compatible with the logical distribution of two-qubit gates. Hence, this results in an extensive application of SWAP operations, especially within the heavy-hex lattice, to accommodate the physical constraints of qubit interactions. Here, the variance in SWAP gate application is directly attributed to the distinct characteristics of each processor topology. Generally, the logical structure of the input circuit, as well as the compilation process features, also affect the SWAP gate count.

\begin{figure*}
\centering
\begin{subfigure}{.33\textwidth}
  \centering
  \includegraphics[width=1\textwidth]{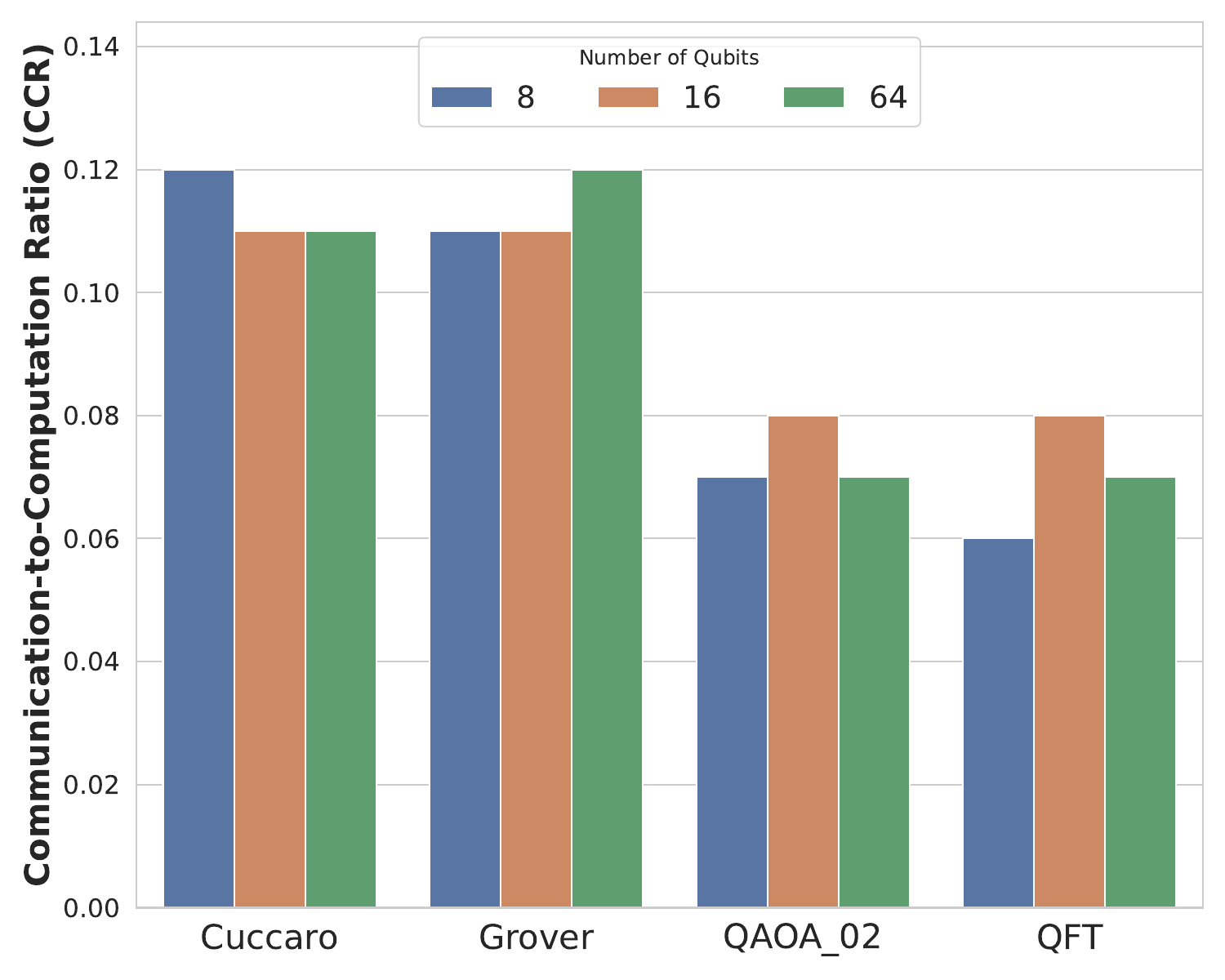}
  \caption{}
\end{subfigure}%
\begin{subfigure}{.33\textwidth}
  \centering
  \includegraphics[width=1\textwidth]{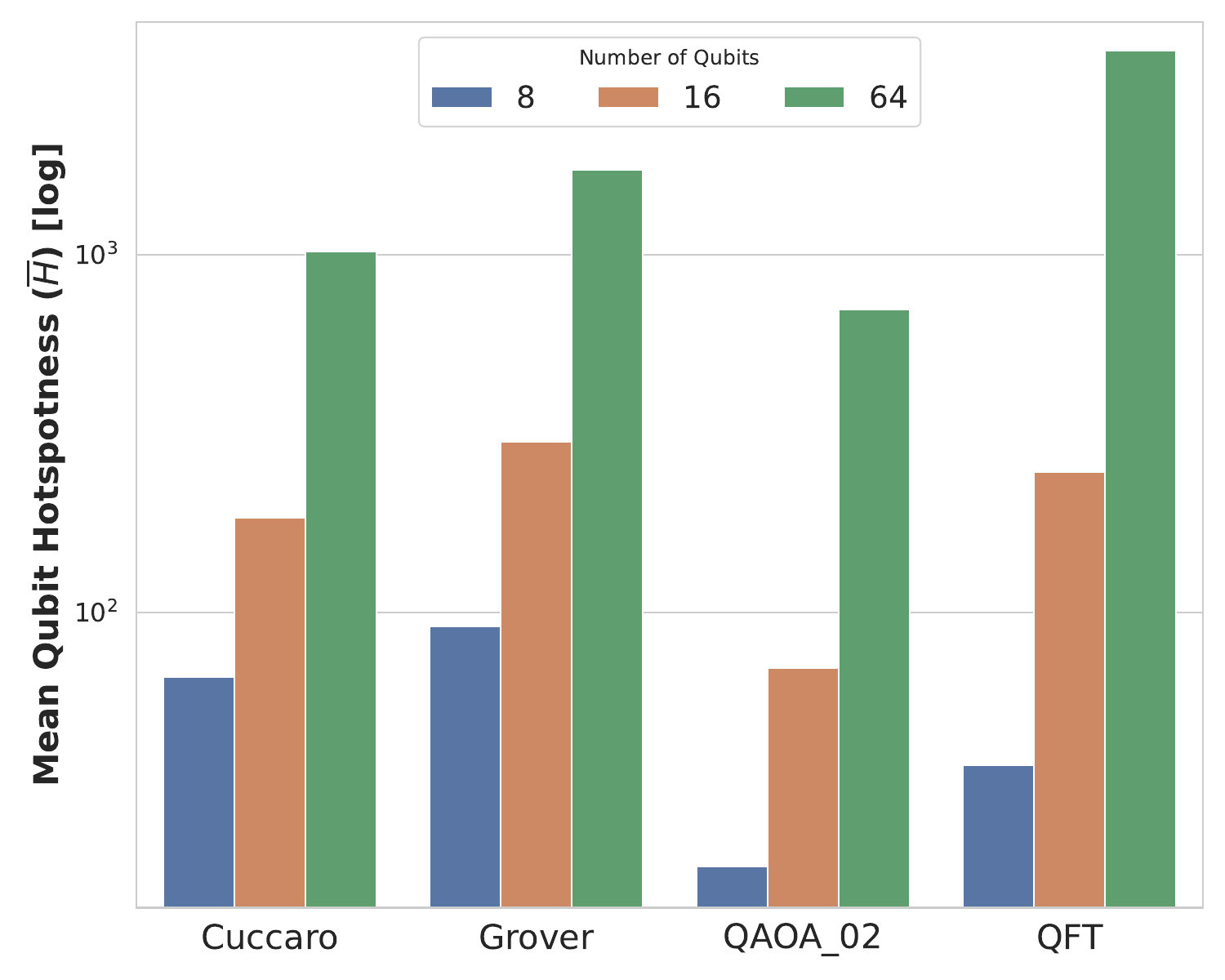}
  \caption{}
\end{subfigure}%
\begin{subfigure}{.33\textwidth}
  \centering
  \includegraphics[width=1\textwidth]{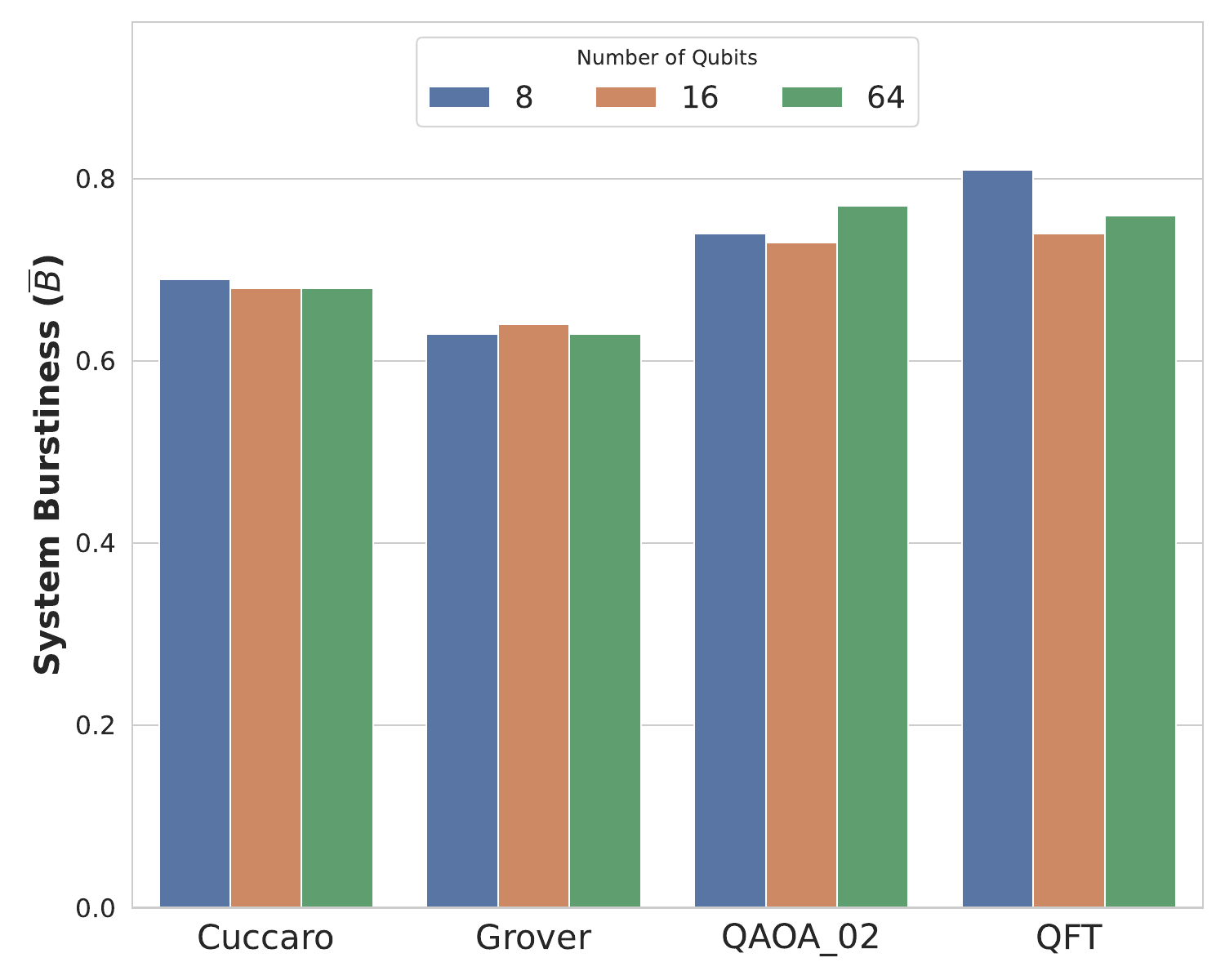}
  \caption{}
\end{subfigure}%
\caption{Communication overhead metrics for circuits compiled to a star topology}\label{star-pm}
\end{figure*}

Within the star topology, our analysis represented in Fig. \ref{star-pm} denotes a communication-to-computation ratio that remains low and stable as we increase the circuit size. The primary factor contributing to this ratio is the central qubit's immediate connectivity to all others, enabling fast state exchanges across the qubits with minimal SWAP operations. However, this same feature leads to an exponential surge in mean qubit hotspotness as the circuit size increases, owing to the significant communication and computation load funneled through the central qubit. Despite this, the low temporal burstiness observed suggests a uniform distribution of SWAP gate execution throughout the computational process, conveying the efficient compiler orchestration of SWAP operations over runtime.

\begin{figure*}
\centering
\begin{subfigure}{.33\textwidth}
  \centering
  \includegraphics[width=1\textwidth]{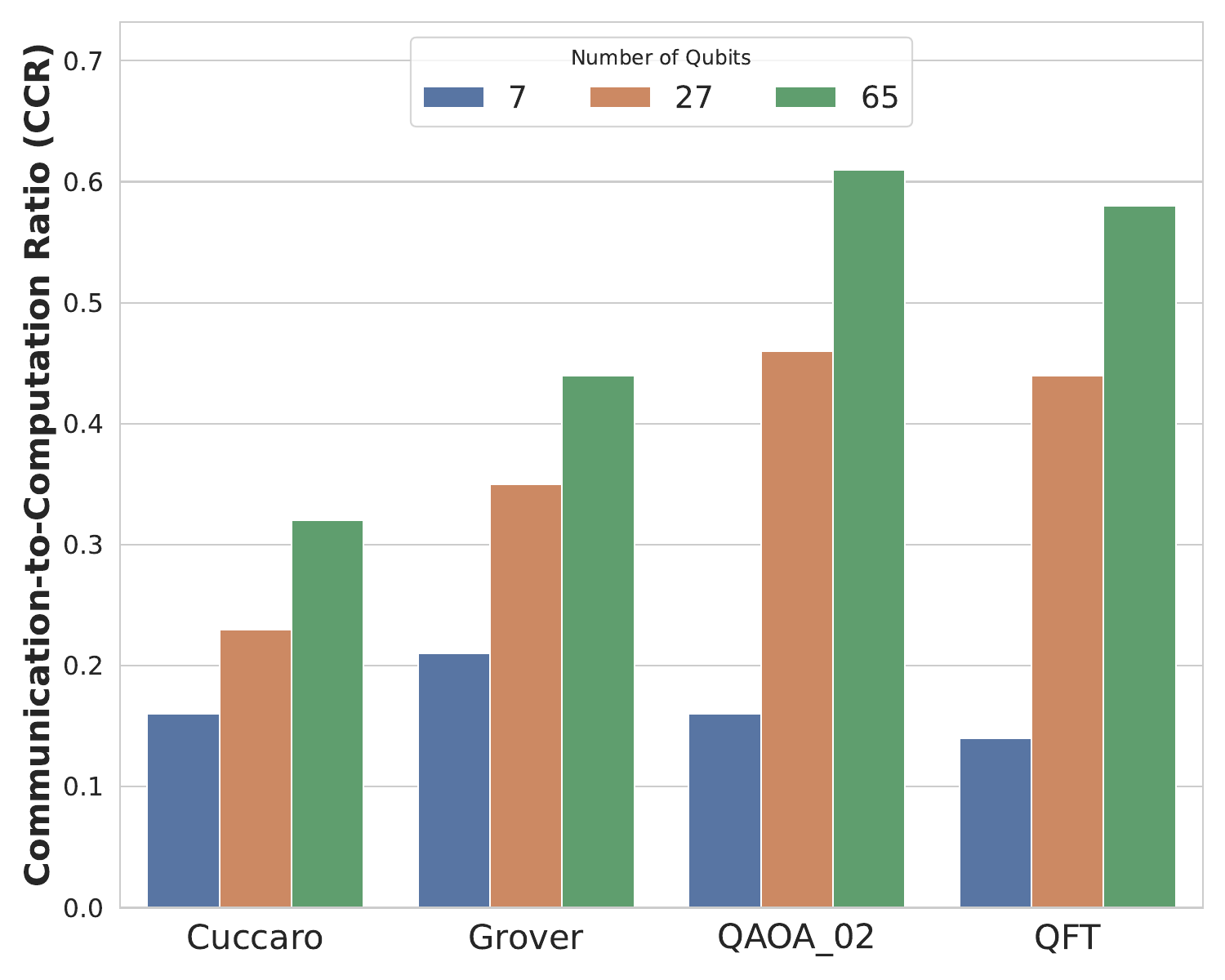}
  \caption{}
\end{subfigure}%
\begin{subfigure}{.33\textwidth}
  \centering
  \includegraphics[width=1\textwidth]{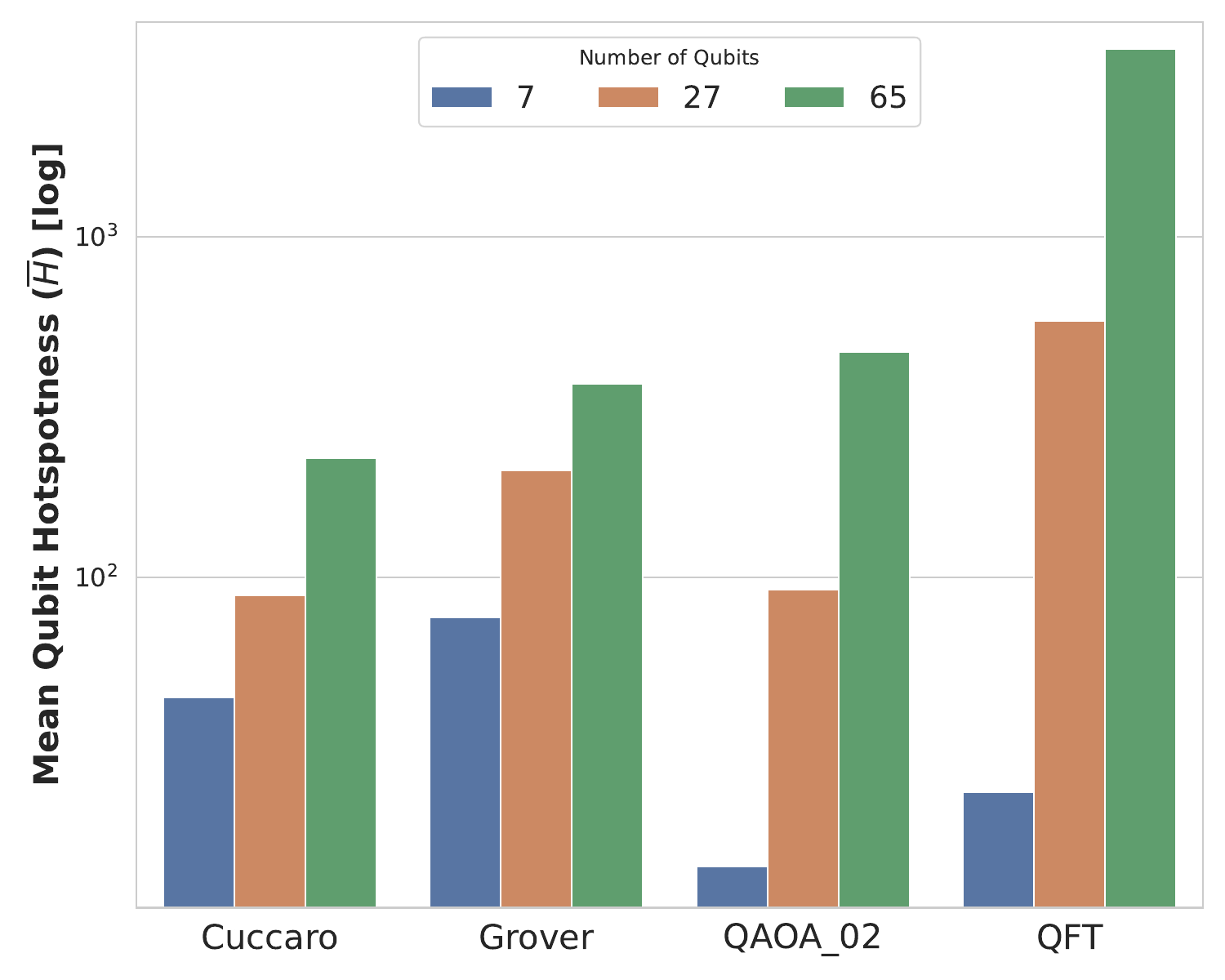}
  \caption{}
\end{subfigure}%
\begin{subfigure}{.33\textwidth}
  \centering
  \includegraphics[width=1\textwidth]{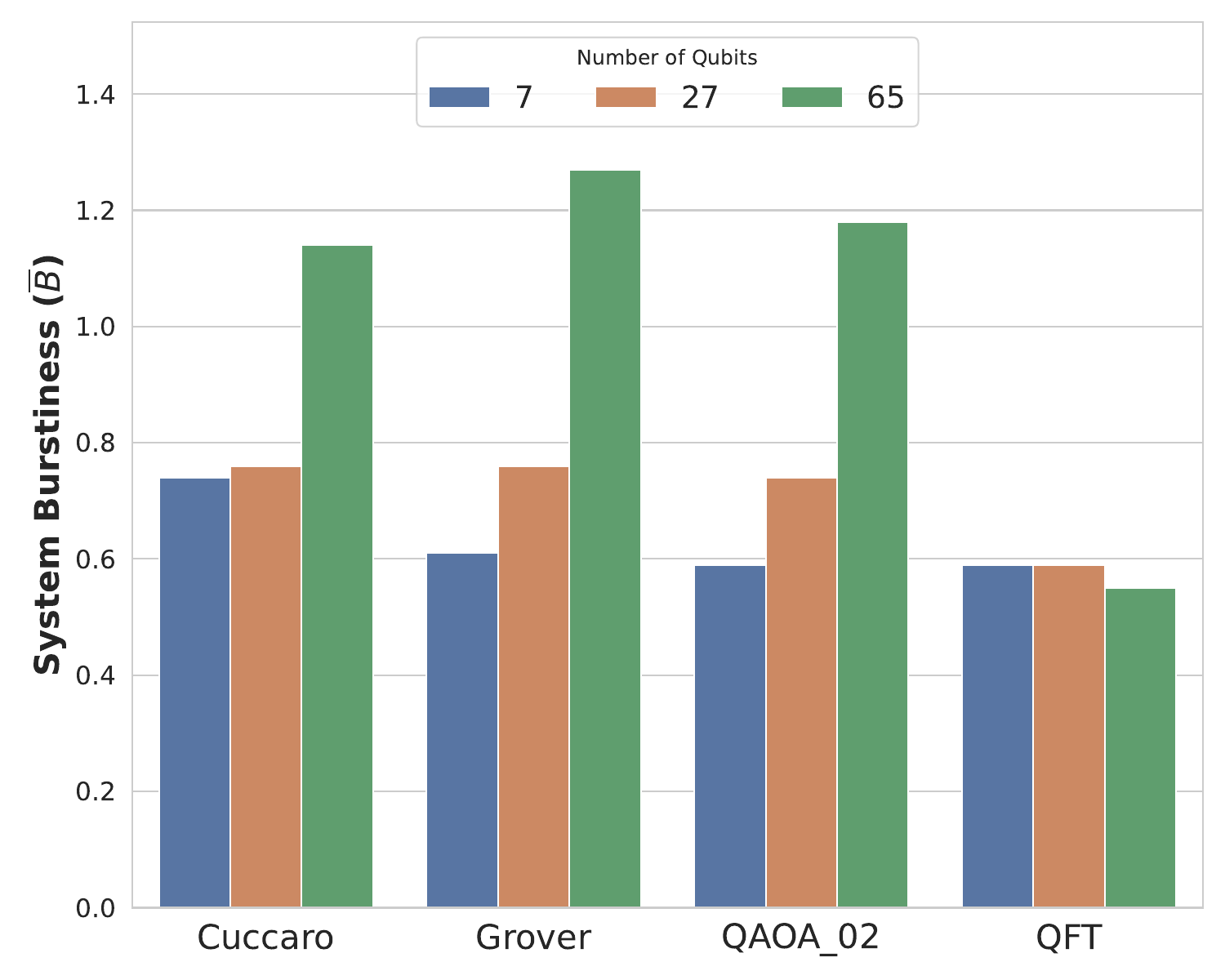}
  \caption{}
\end{subfigure}%
\caption{Communication overhead metrics for circuits compiled to a heavy-hex lattice topology}\label{heavy-hex-pm}
\end{figure*}

As for the heavy-hex topology, which analysis is depicted in Fig. \ref{heavy-hex-pm}, the communication-to-computation ratio maintains a low rate for small circuits, but significantly increases with the circuit complexity. In particular, the communication-to-computation ratio is low in the Cuccaro adder circuit, which displays a cascade of multi-qubit gate applications between neighboring qubits. However, the ratio surpasses $0.5$ in 65-qubit instances of QFT and QAOA\_02, signalling a restrictive communication overhead. The mean qubit hotspotness remains low across most circuits but peaks notably in the 64-qubit QFT instance, where the application of two-qubit gates between distant qubits is prevalent. Temporal burstiness exhibits a significant rise as circuits' sizes scale up specifically for the Grover and Cuccaro programs, indicating the increased necessity to evenly distribute SWAP operations over runtime for similar algorithms.

\begin{figure*}
\centering
\begin{subfigure}{.33\textwidth}
  \centering
  \includegraphics[width=1\textwidth]{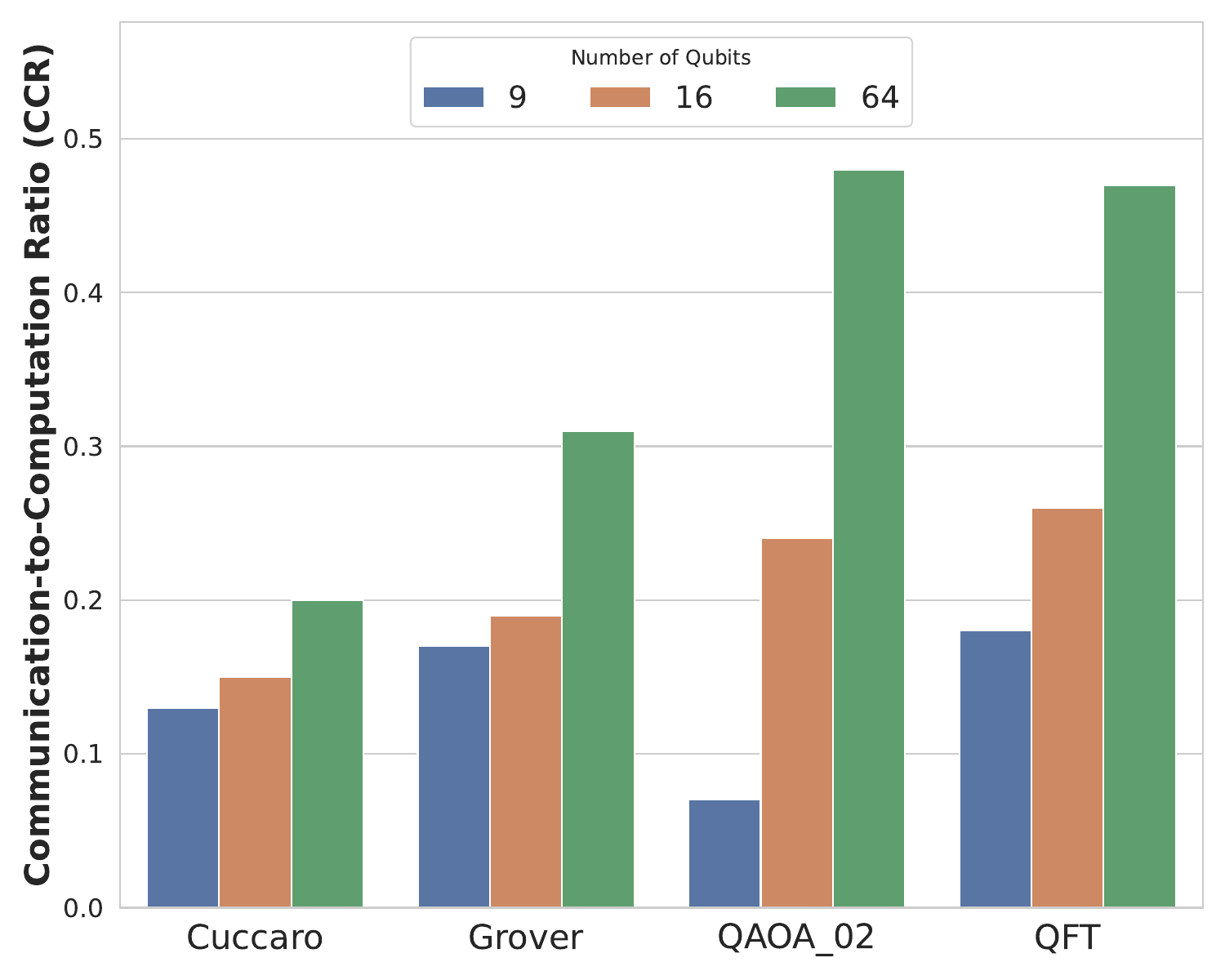}
  \caption{}
\end{subfigure}%
\begin{subfigure}{.33\textwidth}
  \centering
  \includegraphics[width=1\textwidth]{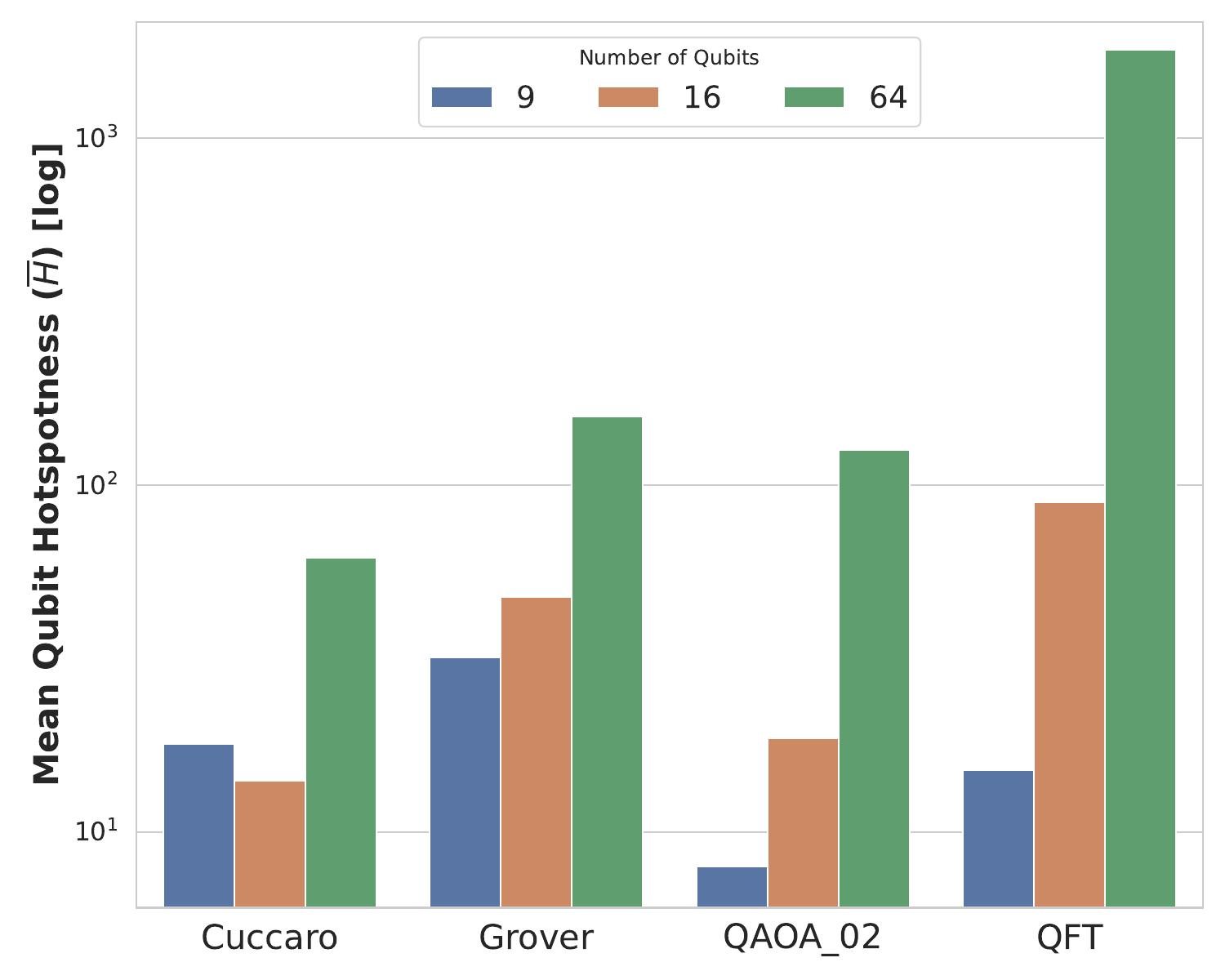}
  \caption{}
\end{subfigure}%
\begin{subfigure}{.33\textwidth}
  \centering
  \includegraphics[width=1\textwidth]{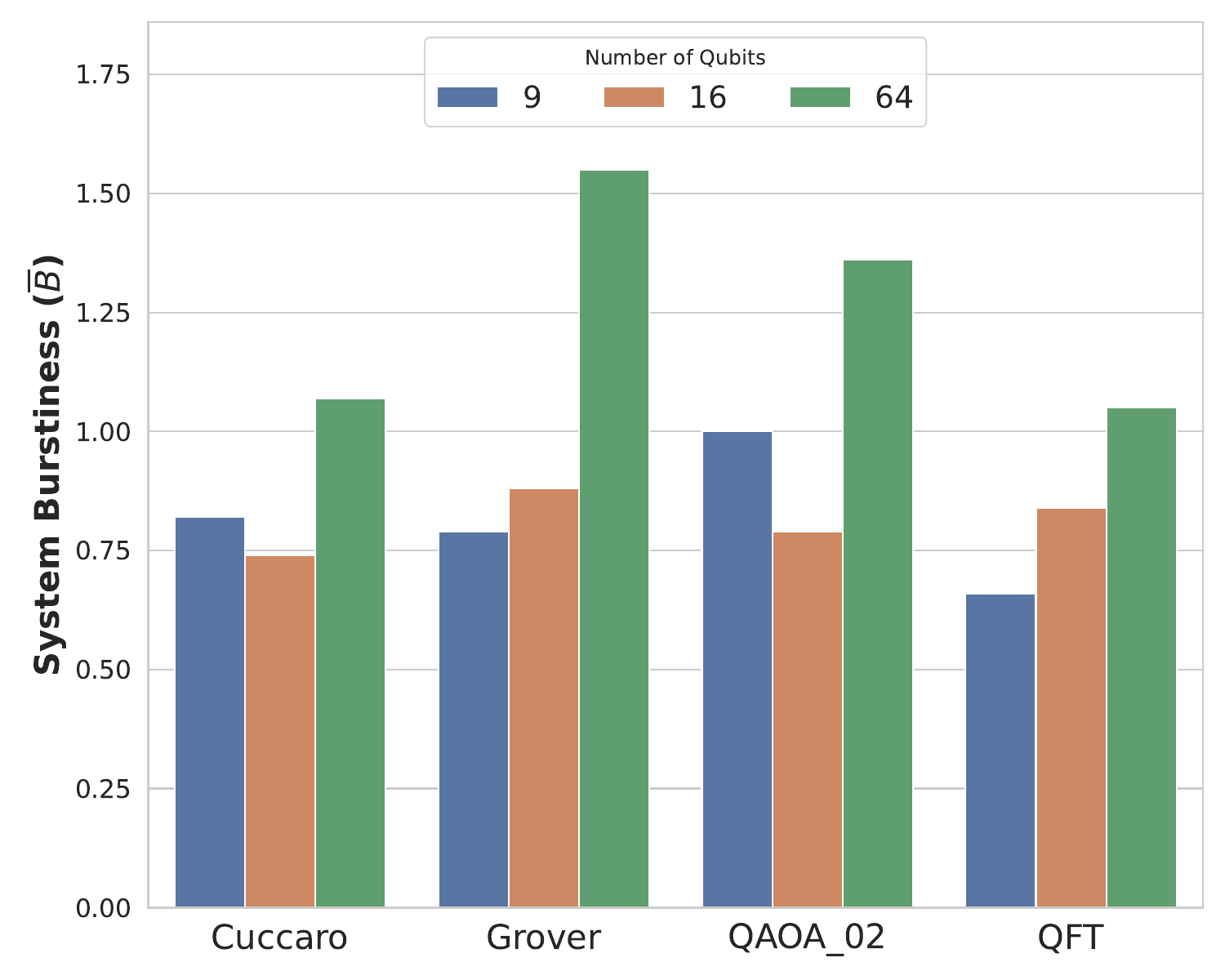}
  \caption{}
\end{subfigure}%
\caption{Communication overhead metrics for circuits compiled to a square lattice topology}\label{square-pm}
\end{figure*}

We present the communication overhead analysis of the square lattice topology in Fig. \ref{square-pm}. This layout, characterized by its higher qubit connectivity compared to the star and heavy-hex lattice, demonstrates a consistently practical communication-to-computation ratio across various circuit sizes. This topology also achieves a lower mean qubit hotspotness, benefiting from the reduced requirement for SWAP gate insertion due to the close spatial arrangement of qubits. The system temporal burstiness is observed to be relatively higher, yet still viable, suggesting a balanced allocation of SWAP gates during program execution.
In principle, while all topologies exhibit a reasonable communication overhead for the examined algorithms and respective circuit sizes, the degree of qubit interconnectivity intrinsic to each topology significantly influences the increase rate and projects the architectural bottlenecks for certain instances. 

\section{Discussion}

We present a comparative analysis of the implications of quantum circuit mapping across the various processor topologies based on a spatio-temporal characterization of the communication overhead originating from the added SWAP gates in qubit routing as imposed by the compiler. The Grover circuit serves as an example to illustrate the circuit structural changes imposed by the compiler to comply with the underlying architecture, offering insight into the distribution of SWAP operations, the frequency of quantum state transfer over runtime, and the involved qubits.

We assess the communication overhead in the star topology. While this topology maintains a low communication-to-computation ratio even as circuit sizes increase, the central qubit becomes a critical bottleneck, reflected in the high mean qubit hotspotness. The concentration of gate applications at the central qubit results in significant traffic that potentially penalizes the computational process with increasing circuit complexity. Additionally, this type of topology limits the paralellization of two-qubit gates since any two-qubit gate has to involve the central qubit, leading to longer execution times of the algorithms. The temporal burstiness remains low, suggesting that the SWAP operations are uniformly distributed throughout the execution time. This efficiency is presumably attributed to the star topology's capacity to enable fast state exchanges, but it does come at the cost of placing substantial demand on the central qubit's connectivity.

The heavy-hex topology, adopted in superconducting qubit systems, demonstrates a nuanced balance between qubit connectivity and circuit mapping complexity. With a generally low communication-to-computation ratio for small-sized circuits, it demonstrates that limited connectivity does not necessarily impede efficient circuit execution. However, the performance of the topology is challenged as the circuits scale and two-qubit gate density increases, as evidenced by the higher communication-to-computation ratio and mean qubit hotspotness for the 64-qubit QFT circuit, predicting a prohibitive communication overhead. The rapid increase in temporal burstiness with circuit size for most instances further illustrates that while this architecture is capable of handling smaller circuits adeptly, larger and more complex circuits may require more sophisticated compilation strategies to manage SWAP gate distribution effectively.

The square lattice topology, a typical 2D topology for quantum processors, offers a higher degree of qubit connectivity, which directly translates into a reduced communication overhead for all examined circuit instances. The mean qubit hotspotness is considerably lower than in the other two layouts, implying a more equitable distribution of communication traffic across the qubits. This characteristic alleviates the central bottleneck issue present in the star topology, and the higher demand for quantum state transfer in sparse architectures such as the heavy-hex, therefore allowing for an execution flow with fewer SWAP gates. This contributes to the observed low temporal burstiness. The square lattice thereby emerges as a compelling topology for quantum processors that is flexible with the various quantum algorithms, and it appears to scale more efficiently with increasing circuit complexity.

\section{Conclusion}
We emphasize the intricate dynamics of the communication overhead, structural aspects of quantum circuits and the configuration of quantum processors, and evaluate the impact of the compilation process from a communication perspective. We show that the square lattice topology, characterized by a robust performance across all metrics, remains a favourable and versatile layout for scaling quantum processors compared to the studied architectures. In future work, we analyze the communication overhead in modular quantum computers architectures according to a larger set of performance metrics.

\end{document}